# A Zonal Volt/VAR Control Mechanism for High PV Penetration Distribution Systems

Asmaa Alrushoud, Catie McEntee, and Ning Lu

Department of Electrical and Computer Engineering

North Carolina State University

Raleigh NC, USA

aalrush@ncsu.edu

*Abstract*—**This paper presents a zonal Volt/VAR control scheme for regulating voltage in unbalanced 3-phase distribution systems using inverter-based resources (IBR). First, the dependency between nodal voltage changes and IBR reactive power injections is derived via voltage sensitivity studies. Then, a fast-incremental clustering method is used to divide the distribution circuit into weakly-coupled zones based on correlations between nodal voltage sensitivities. The weak-coupling allows the voltage to be regulated independently within each zone using a rule-based voltage controller to dispatch IBR for voltage corrections. Simulation results on actual distribution feeder show that the proposed zone-based Volt/VAR control method maintains system voltages within their operational limits while reducing the runtime of the Volt/VAR controller from tens of second to a couple of milliseconds compared with centralized, optimization-based Volt/VAR control methods.**

## I. INTRODUCTION

Voltage regulation (VR) methods using distributed energy resources (DERs) are categorized into centralized [1], local [2] and distributed [3]. Centralized methods, formulated as global optimization problems, normally require full-network observability and rely on high-bandwidth communication infrastructures for monitoring and control. Local control strategies, aiming at regulating the voltage at a specific node, require only local measurements. Distributed control methods, allowing coordination among devices, employ low-bandwidth communication links for a limited amount of data exchange between units to achieve a joint control objective at the least cost using all resources available at the time. The zonal VR algorithm propose in this paper belongs to the distributed approach. A large distribution network is partitioned into non-overlapping, weakly-coupled VR zones based on correlations between nodal voltage sensitivities. This allows voltage control of the entire network to be effectively achieved by controlling the voltage inside each VR zone.

Methods for partitioning distribution network have been reported in [4-9]. In [4], Biserica *et al.* proposed a zoning method based on k-means clustering algorithm for ancillary services in distribution networks with distributed generation. In [5], Ding *et al.* divided the distribution network into several sub-communities based on the spectral clustering algorithm to provide voltage control. An affinity propagation clustering algorithm is employed in [6] by Li *et al.* to divide the distribution network and proposes a distributed Volt/VAR control method to minimize power loss. In [7], Zhao *et al.*

proposed a network partition method based on a community detection algorithm to perform zonal voltage control. Also, particle swarm optimization (PSO) is applied to achieve optimal control by minimizing the amount of active/reactive power control in clusters. Network partitioning algorithm based on binary particle swarm optimization (BPSO) was presented by Li *et al* in [8] to provide a two-stage decentralized optimal reactive power dispatch. A power network partition method based on power flow tracing and agglomerative algorithm for reactive power control is presented by Gong *et al* in [9].

The aforementioned approaches, when applying clustering algorithms to achieve network partition, requires a distance metric for describing the relationships among nodes in the distribution network. The distance metric is defined by a modified electrical distance based on the impedance distance, reactive power-voltage magnitude sensitivity, and modularity index in [5], [6, 8, 9], and [7], respectively. However, all those approaches use optimization-based approaches for zonal voltage control. The computational cost for minimizing active power losses or voltage fluctuations within each VR zone is high. In addition, the algorithms were developed and tested only on very small IEEE test systems [5-9] or/and small balanced distribution systems [4, 5, 7], making it hard to scale up to large, unbalanced distribution feeders with hundreds or thousands of load nodes that supply a mix of 1-phase, 2-phase, and 3-phase unbalanced loads.

In this paper, we propose a rule-based, distributed VR scheme to reduce the runtime to milliseconds without the need for sophisticated compute engines. This enables VR control at grid-edge on a full-size utility feeder possible. The contributions of the paper are summarized as follows. *First*, we developed a VR-zone identification methodology for dividing load nodes on a distribution feeder into VR zones so that each VR zone consists of a collection of nodes with strongly correlated voltage-load sensitivities whereas nodes in different zones are weakly correlated. *Second*, based on the voltage-load sensitivity, a distributed, priority-list based zonal VR control scheme is developed to correct voltage violations in its local zone as a non-optimization based VR control. This greatly simplified the control algorithm and shortened the computation speed. *Third*, the unbalanced nature of distribution feeders is accounted for by treating each phase separately.

A centralized control presented in [10] is used as a benchmark for comparing the performance of the proposed zonal-based control on an actual full-size utility feeder. Simulation results demonstrate that the proposed algorithm significantly reduces the running time while achieving comparable performance with the centralized optimization-based control approach.

## II. Zone-based Voltage Regulation Method

This section presents the zone-based VR regulation method.

### A. Voltage-Load Sensitivity Analysis

In this paper, we use the power flow based voltage-load sensitivity matrix (VLSM) calculation method introduced in [11] to calculate the VLSM with respect to reactive power changes, $VLSM_Q$, expressed as

$$VLSM_Q = \begin{bmatrix} q_{1,1} & q_{1,2} & \cdots & q_{1,N} \\ q_{2,1} & q_{2,2} & \cdots & q_{2,N} \\ \vdots & \vdots & \ddots & \vdots \\ q_{N,1} & q_{N,2} & \cdots & q_{N,N} \end{bmatrix} \quad (1)$$

The $j^{th}$ column of $VLSM_Q$ is calculated as

$$VLSM_Q(:,j) = \begin{bmatrix} q_{1,j} \\ \vdots \\ q_{N,j} \end{bmatrix} = \frac{V_{i,j}^Q - V_i^{BASE}}{\Delta Q_j} \quad \forall i \in N, \forall j \in N \quad (2)$$

where $N$ is the number of load nodes; $V_i^{BASE}$ is the base voltage at load node $i$ for a given power flow case; $\Delta Q_j$ is the reactive power perturbation at node $j$; $V_{i,j}^Q$ is the voltage at node $i$ when perturbing node $j$ by $\Delta Q_j$. It should be noted that $N$ represents the number of single-phase load nodes on a feeder, i.e., one 3-phase load node is represented as three single-phase load nodes.

### B. Clustering Method for VR Zone Identification

Distribution feeders are inherently unbalanced because line transposition is usually not used and phase loading is always fluctuating [12]. Consequently, all three phases need to be represented and treated differently. Therefore, we group loads on the same phase first before clustering, as shown in Fig. 1.

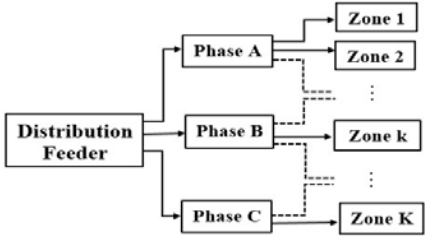

Fig. 1. Distribution feeder VR zone clustering process.

In principle, a VR zone consists of load nodes with highly-correlated nodal voltage changes. To determine the number of VR zones in a feeder and identify the number of load nodes in each VR zone, we apply the fast incremental clustering (FIC) method for clustering load nodes using $VLSM_Q$ computed at different loading conditions. FIC is a non-iterative algorithm [13] for classifying data one-by-one in a sequential manner. Thus, each load node is considered only once and after the classification of a load node is completed, its membership to a cluster will not change.

Because FIC relies on the order in which the data is processed, we first sort the load nodes in ascending order according to their distance to the substation before FIC is applied. This sorting is necessary because voltage sensitivity increases with distance.

At each time interval $t$, compute $VLSM_Q$ and calculate the mean and the standard deviation of the $j^{th}$ column of $VLSM_Q$, as $\mu_j$ and $\sigma_j$. Then, compute the voltage sensitivity correlation matrix, $\mathbf{C_Q}$ as

$$\mathbf{C_Q} = \begin{bmatrix} 1 & c_{1,2} & \cdots & c_{1,N} \\ c_{2,1} & 1 & \cdots & c_{2,N} \\ \vdots & \vdots & \ddots & \vdots \\ c_{N,1} & c_{N,2} & \cdots & 1 \end{bmatrix} \quad (3)$$

$$c_{m,n} = \frac{1}{N} \sum_{i=1}^N \left( \frac{q_{i,m} - \mu_m}{\sigma_m} \right) \left( \frac{q_{i,n} - \mu_n}{\sigma_n} \right) \quad \forall m \in N, \forall n \in N \quad (4)$$

where $c_{m,n}$ is the Pearson correlation coefficient [13] for evaluating the correlation between any two columns, $m$ and $n$, in $VLSM_Q$. The values of the correlation coefficients $c_{m,n}$ range between -1 and 1, where "-1" represents a negative correlation, "0" represents no correlation, and "1" represents a positive correlation [13].

The mean voltage sensitivity correlation coefficient between a new load node $l$ and the $m_k$ load nodes in an existing VR zone $k$, $mcc_{l,k}$ is calculated as

$$mcc_{l,k} = \frac{1}{m_k} \cdot \sum_{x=1}^{m_k} c_{l,x} \quad \forall k \in K \quad (5)$$

where x is the index of the load node inside VR zone $k$ and $K$ is the total number of VR zones.

If $mcc_{l,k}$ calculated for all $K$ zones are below the correlation threshold, $\alpha$, a new VR zone will be created. Otherwise, put node $l$ into an existing VR zone with the maximum $mcc_{l,k}$. Because FIC groups load nodes by comparing $mcc_{l,k}$ to $\alpha$, increasing $\alpha$ will increase the number of VR zones. The process of FIC is summarized in Algorithm 1.

| Algorithm 1: Fast Incremental Clustering |
|---|
| 1:    Set the number of VR zones to be $K = 1$ and select a correlation threshold, $\alpha$. |
| 2:    Place node 1 into VR zone 1. |
| 3:    Compute the correlation matrix $C_Q$ using (3). |
| 4:    **for** node $l=2...N$ |
| 5:       Use (5) to calculate $mcc_{l,k}$ for all $K$ VR zones. |
| 6:       **if** $\max_{k \in K} (mcc_{l,k}) \geq \alpha$ |
| 7:         Place $l$ in the zone having maximum $mcc_{l,k}$. |
| 8:       **else** |
| 9:         $K = K + 1$ |
| 10:      Place $l$ in the new VR zone $K$. |
| 11:      **end** |
| 12:      $l = l + 1$ |
| 13:   **end** |

### C. Zone-based Volt/VAR Control Mechanism

After all $N$ nodes on a feeder are grouped into $K$ zones using the method developed in Section II-B, for voltage violations inside a VR zone, we can use VR resources inside that zone to alleviate or remove those violations because all nodes inside a VR zone are highly-correlated. This will also allow us to prioritize VR resources inside a VR zone based on $c_{m,n}$, which is an indicator for how effective the nodal reactive power injection at node $m$ is for regulating the voltage on node $n$. The priority-list based dispatch is straightforward and computationally efficient comparing to the optimization-based VR resource dispatch.

At every time step $t$, if any voltage violations are detected in VR zone k, the $k^{th}$ VR zone controller will be in voltage corrective mode and will identify load node $r$ with the highest over-voltage, $V_{r,k}^+(t)$, or the lowest under-voltage, $V_{r,k}^-(t)$ and compute its voltage violation, $\Delta V_r(t)$, as

$$\Delta V_r(t) = \begin{cases} V_{r,k}^+(t) - V_{max} & s.t. \ V_r^+(t) = \max_{x \in m_k} V_x \\ V_{min} - V_{r,k}^-(t) & s.t. \ V_r^-(t) = \min_{x \in m_k} V_x \end{cases} \quad (6)$$

where $V_{max}$ and $V_{min}$ is the predetermined nodal voltage upper and lower limit, respectively.

Then, the required reactive power, $\Delta Q_{Req}(t)$, for correcting $\Delta V_r(t)$ can be calculated as

$$\Delta Q_{Req}(t) = \frac{\Delta V_r(t)}{\alpha \cdot q_{r,r}} \quad (7)$$

where $q_{r,r}$ is the voltage-reactive power sensitivity at node $r$ from the $VLSM_Q$. Because all the load nodes within a zone are highly correlated at $\alpha$ correlation level, the correction of the largest nodal voltage violation will let all other nodal voltages inside the VR zone return to the normal range. Note that when $\Delta Q_{Req}(t)$ is positive/negative, the system requires reactive power injection/absorption.

To regulate the voltage at node $r$ in the $k^{th}$ VR zone, all node $x$ in VR zone k will be assigned a priority based on $c_{x,r}$ ($\forall x \in m_k$) calculated by (4) in the descending order. Thus, a distributed, priority-list based zonal VR control scheme is developed. Let $PL_k$ represent the priority list in the $k^{th}$ VR zone, which consists of a set of $VLSM_Q$ ranked nodes indexed by $p_k$, where $p_k \in [1 \ m_k]$. The controller dispatch from the node on the top of the priority list in a sequential manner until $\Delta Q_{Req}(t)$ is met or all the resources within the zone are exhausted. If the node does not contain a smart inverter, the controller would go to the next node on the priority list.

If voltage violations are no longer detected, the controller will be in return mode where it controls the smart inverters $D_k$ within each zone to return to the unity power factor mode based on the reversed priority list. This means the inverters at the end of the priority list will return to unity power factor first. To avoid voltage oscillations caused by the sudden drop of reactive power supply, a voltage return band, $[(V_{min} + \varepsilon_d), (V_{max} - \varepsilon_u)]$, is defined, where $\varepsilon_u$ and $\varepsilon_d$ are two voltage return margins. Thus, when $V_{r,k}^+(t)$ and $V_{r,k}^-(t)$ are within the voltage return band, we will use (6) and (7) to calculate $\Delta Q_{Req}(t)$, which guide the dispatch of the smart inverters $D_k$ back to their unity power factor operation mode based on the priority list.

The zonal Volt/VAR control algorithm is summarized in Algorithm 2.

---

**Algorithm 2: Zone-based Volt/VAR Control**

1:     At time $t$, run power flow simulation. Set $p_k = 0, D_k = m_k$.
2:     **for** $k = 1$: $K$
3:        **if** $V_{max} \geq V_{r,k}^+ \geq V_{max} - \varepsilon_u$ or $V_{min} + \varepsilon_d \geq V_{r,k}^- \geq V_{min}$
4:           The $k^{th}$VR zone controller is in idle mode.
5:        **else if** $V_{r,k}^+ > V_{max}$ or $V_{r,k}^- < V_{min}$
6:           The $k^{th}$VR zone controller is in voltage corrective mode.
7:        **else if** $V_{r,k}^+ < V_{max} - \varepsilon_u$ or $V_{r,k}^- > V_{min} + \varepsilon_d$
8:           The $k^{th}$VR zone controller is in return mode.
9:        **end if.**
10:     Calculate $\Delta V_r(t)$ and $\Delta Q_{Req}(t)$ using (6) and (7)
11:     Rank all nodes based on $c_{x,r}$ in descending order to obtain $PL_k$
12:     **while** $\Delta Q_{Req} > 0$
13:        **if** in corrective mode and $p_k < m_k$
14:           $p_k = p_k + 1$
15:           Select node $PL_k(p_k)$
16:           **if** this is a PV node
17:              $Q_{PL(p_k)} = Q_{PV,rated}$, $flag(PL_k(p_k)) = 1$
18:           **else** $Q_{PL(p_k)} = 0$
19:           **end if**
20:           Update $\Delta Q_{Req} = \Delta Q_{Req} - Q_{PL(p_k)}$
21:        **else if** in corrective mode and $p_k \geq m_k$
22:           Resources exhausted so $p_k = m_k$ and let the $k^{th}$ VR zone controller idle
23:        **else if** in return mode
24:           **if** $flag(PL_k(D_k)) = 1$
25:              Update $\Delta Q_{Req} = \Delta Q_{Req} - Q_{PL(D_k)}$
26:              $Q_{PL(D_k)} = 0, flag(PL_k(D_k)) = 0$
27:           **end if**
28:           $D_k = D_k - 1$
29:        **else if** in return mode and $D_k = 0$
30:           Let $k^{th}$VR zone controller idle.
31:        **end if**
32:     **end while**
33:     **end for**

---

### D. Benchmark Case

We use a centralized optimal Volt/Var control (CVVC) method introduced in [10] as a benchmark case for comparing the performance of the zonal Volt/Var control method. The CVVC controller requires full visibility of the real-time voltage at each node and the real-time $VLSM_Q$ re-calculated every 5 minutes using the same procedure described in section II.A.

The CVVC optimization is formulated as

$$\min_{d \in D}(\Sigma \ C_d(Q_d^+, Q_d^-)) \quad (9)$$

$$Q_{d,min} < Q_d^+ - Q_d^- < Q_{d,max} \ \forall d \in D \quad (10)$$

$$V_{min} < V_j + \Sigma_{d \in D}(\Delta Q_d \times \rho_{Qj,d}) < V_{max} \ \ \forall j \in N \quad (11)$$

$$\Delta Q_d = Q_d^+ - Q_d^- - Q_d^0 \quad (12)$$

$$C_d(Q_d^+, Q_d^-) = c_d \times (Q_d^+ + Q_d^-) \quad \forall i \in N \quad (13)$$

where $Q_d^+$ and $Q_d^-$ are the decision variables represent injection and absorption of reactive power for each inverter $d$ in the entire set of inverters on the circuit, $D$.

Reactive power is limited in (10) to the remaining inverter capacity after considering the active power output and (11) enforces voltage constraints. The change in reactive power $\Delta Q_d$ is the difference between the new net reactive power injection and the initial reactive power injection $Q_d^0$ at the beginning of the control cycle which was used in the initial power flow, as shown in (12). The cost for reactive power represented by (13) is a flat cost per kvar.

### III. SIMULATION RESULTS

A three-phase unbalanced distribution feeder model representing a real circuit located in North Carolina is selected to verify the effectiveness of the proposed methods. The feeder includes 1388 nodes, 415 of which are load nodes. The feeder load disaggregation algorithm presented in [14] is used to allocate 1-min resolution weekly residential load profiles from Pecan Street [15] to every load node on a test feeder. This step is important to give each load node a realistic individual load profile and will prepare the nodal load profiles for running quasi-static power flow simulations. The disaggregation algorithm allocates 1024 households to the feeder with a feeder load peak at 3.52 MW.

To model high-PV penetration scenarios, we assume that each household has a 5 kW PV system connected to it. At the feeder level, the PV installation is 5.12 MW. We assume that each inverter can inject and absorb reactive power and is sized such that its total apparent power $S$ can reach up to 110% of its rated active power capacity $P$.

The maximum real-time PV penetration situation is considered as the most critical scenario because it withstands the most severe voltage variations as seen in Fig. 2. Thus, we will use the $VLSM_Q$ results at the highest PV penetration and it will not be updated through the weekly simulation. Note that the PV penetration is defined as the percentage of the peak of the aggregated PV output to the peak load on the feeder.

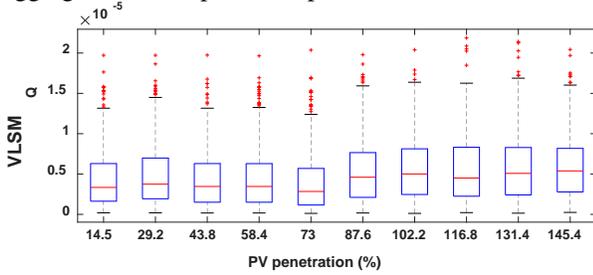

Fig. 2. $VLSM_Q$ for different PV penetrations scenarios.

### A. VR Zone Identification

We perform FIC at different PV and load levels for different operation conditions until a stable number of clusters are generated. In Fig. 3, we evaluate the impact of different values of $\alpha$ on the number of VR zones identified in each phase and have the following observations:

- For each phase, the number of VR zones can be different, demonstrating the needs for treating each phase separately. This can also be demonstrate by the fact that if $\alpha \leq 0.92$, all nodes on a phase can be clustered into a single VR zone.
- When $\alpha$ increases, the number of VR zones increases. Increasing the number of VR zones will increase the number of zonal VR controllers but will achieve higher voltage variations correlation level within each zone.

We will be comparing zonal voltage control results under two correlation thresholds. At 92%, where each phase is considered as a zone by itself summing to 3 zones, and at 96%, where each phase is divided into two zones summing to 6 zones. The proposed zones boundaries under 92% correlation threshold are shown in Fig. 4.

It should be noted that three-phase load nodes are equipped with three single-phase inverters and in this case each inverter will belong to a different zone according to the phases. So in Fig.4, at the top of the feeder where most of the load nodes are three-phase nodes, each inverter will be belong to a zone according to its phase which cannot be shown graphically.

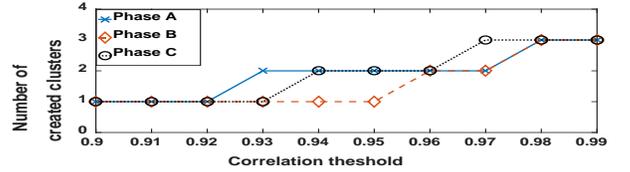

Fig. 3. Impact of $\alpha$ on the number of VR zones.

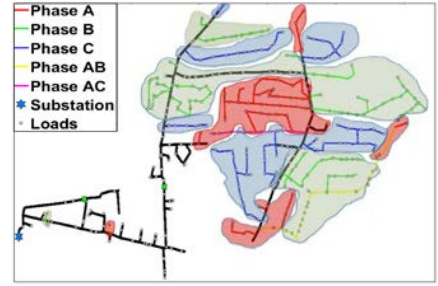

Fig. 4. VR zones identification on a distribution feeder ($\alpha = 92\%$)

### B. Zonal Volt/VAR Control Performance

In this section, two cases (i.e., 3-zone and 6-zone) are used to compare the number of VR zones on the zonal VR controller performance. In both cases, set $V_{min} = 0.951$ p.u. and $V_{max} = 1.049$ p.u., which are within the ANSI limits. As shown in Fig. 5, where each color curve represents voltage profile of a load node, dividing the feeder into 3 or 6 VR zones results in similar performance where both controller regulate the voltage profile within the ANSI limits at similar computational time.

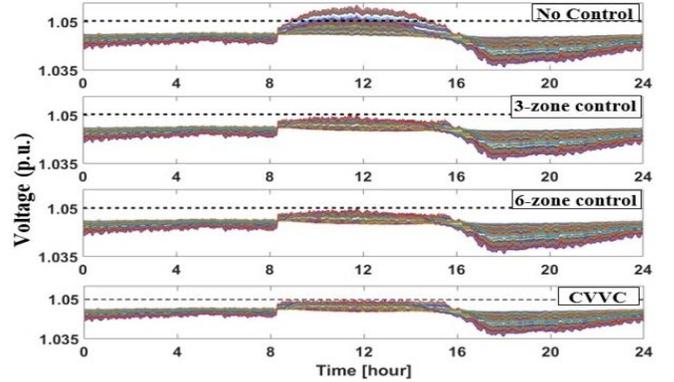

Fig. 5. Comparison of daily voltage profiles.

### C. Comparison with the CVVC algorithm

In Fig. 6, the reactive power absorption needed from phase A smart inverters (from 8 a.m. till 5 p.m.) to correct voltage violations at node 411 Phase A (i.e. $V_{r1}^+$) is shown for zonal and centralized control. It can be seen that the zonal control estimates near optimal reactive power absorption during voltage corrective mode which validates the effectiveness of proposed control scheme. Also, when the controller is in return mode, the zonal control underestimates its Q absorption reduction as compared with the centralized controller and that is expected because there is a return voltage band within the zonal controller logic. A comparison has been carried out for *with* and *without* voltage return margins, where we show that the introduction of the voltage return band can reduce the voltage oscillations but underestimate the required $Q$ reduction. The results also show that $\varepsilon_u = 0.001$ p.u. yields the best performance.

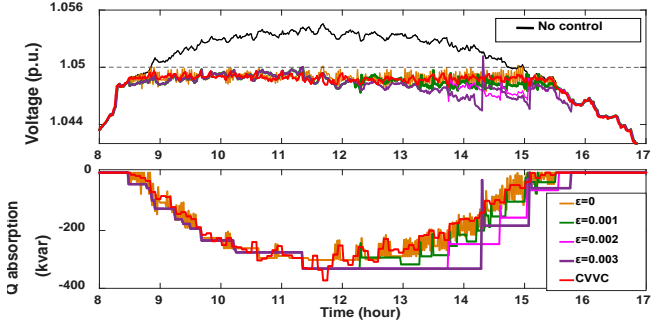

Fig. 6. Voltage profiles and reactive power compensation at Node 411.

In Fig. 7 (a), we show a violin plot for the distribution of the maximum voltage for a whole week for the base case (no control), the zonal control case (3-zone), and the centralized control. The plots show that the zonal voltage control algorithm can maintain the voltage profile within desired limits under time varying condition with comparable performance as that of optimization based methods. In Fig. 7 (b), we show a violin plot for comparing the total reactive power absorption for both zonal and centralized control. The results show that the zone-based, priority-list VR control algorithm is robust and can estimate adequately the required reactive power compensation needed to mitigate voltage issues over extended periods of time without updating the clustering scheme or the $VLSM_Q$.

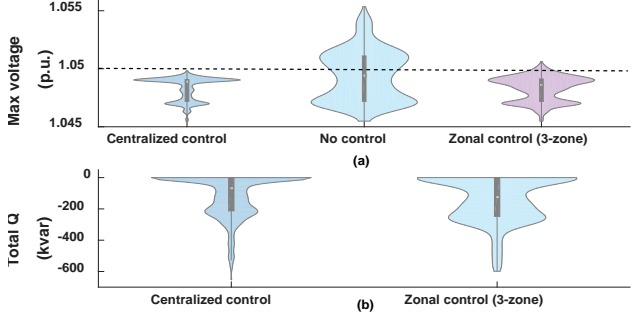

Fig. 7. Comparisons of (a) the maximum voltage and (b) $Q$ absorption.

### D. Computational Speed

In Fig. 8, we compare the runtime of the proposed algorithm with the optimization based methods. For the optimization based centralized control, the boxplot represents the time to calculate the VLSM and to solve the optimization objective function. The runtime for the zone-based priority-list VR control is consistently at approximately 2 milliseconds, while the runtime of the optimization-based approach ranges between 73 to 171 seconds with an average of 138 seconds. The results are obtained using an Intel i7-4790 CPU processor.

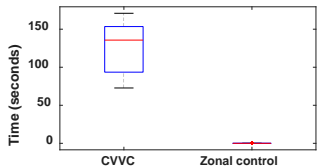

Fig. 8. Runtime comparison between the centralized and zonal VR control.

## IV. CONCLUSION AND FUTURE WORK

In this paper, we presented a zone-based VR control algorithm. First, the feeder load nodes are divided into VR zones per each phase based on voltage reactive power sensitivity. Next, a priority list is obtained in each VR zone by ranking each node according to its VR regulation capability. The priority list enables a non-optimization based VR control approach for correcting over/under voltage violations inside each zone using local smart inverters. Due to the weak coupling among different VR zones, we can effectively locate the most efficient smart inverters for correcting voltage violations without significant altering voltage profiles in other VR zones. The proposed method is tested on an actual distribution feeder. Our results show that it is effective for dividing the distribution feeder into VR zones for maintaining zonal voltage within desired limits using local resources. Comparisons with a centralized-based, optimization based Volt/VAR control show that the zonal Volt/VAR control gives near optimal reactive power compensation. Our future work will be focused on coordination between utility control devices (*i.e.* voltage regulators and capacitors) and cross-zone smart inverters.